\documentclass[letterpaper,twocolumn,10pt]{article}

\pdfoutput=1

\usepackage{graphicx}
\usepackage{cite}
\usepackage{xcolor}
\usepackage{xifthen}
\usepackage{hyperref}
\usepackage{listings}
\usepackage{xcolor}
\usepackage{multirow}

\date{}

\title{\Large\bf Policy/mechanism separation in the Warehouse-Scale OS}

\author{
{\rm Mark Mansi}\\
University of Wisconsin-Madison
\and
{\rm Michael M. Swift}\\
University of Wisconsin-Madison
}

\newcommand{\sys}{Grape CM}

\newcommand{\todo}[1][]{ \ifthenelse{\isempty{#1}}{{\color{red}[TODO]}}{{\color{red}[TODO: #1]}} }

\begin{document}

\maketitle

\begin{abstract}

\textit{``As many of us know from bitter experience, the policies provided in extant operating systems, which are claimed to work well and behave fairly `on the average', often fail to do so in the special cases important to us''}~\cite{hydra}.
Written in 1974, these words motivated moving policy decisions into user-space.
Today, as warehouse-scale computers (WSCs) have become ubiquitous, it is time to move policy decisions away from individual servers altogether.
Built-in policies are complex and often exhibit bad performance at scale.
Meanwhile, the highly-controlled WSC setting presents opportunities to improve performance and predictability.

We propose moving all policy decisions from the OS kernel to the cluster manager (CM), in a new paradigm we call \sys{}. In this design, the role of the kernel is reduced to monitoring, sending metrics to the CM, and executing policy decisions made by the CM.
The CM uses metrics from all kernels across the WSC to make informed policy choices, sending commands back to each kernel in the cluster.
We claim that \sys{} will improve performance, transparency, and simplicity.
Our initial experiments show how the CM can identify the optimal set of huge pages for any workload or improve memcached latency by 15\%.

\end{abstract}

\section{Introduction}

In the early 2000s, service providers realized that building bigger, faster, more fault-tolerant servers was an impractical way to handle more traffic.
They turned instead to large clusters of commodity hardware and general-purpose operating systems.
Today, warehouse-scale computers (WSCs) have become a ubiquitous technique for service providers to operate large-scale services~\cite{barrosoDatacenterComputerIntroduction2013}.
However, while general-purpose OSes have allowed the rise of WSCs, they present challenges and missed opportunities, too.
In particular, their built-in policies largely ignore a WSC's unique combination of relative homogeneity and slow-changing workload mix.
We assert that in a WSC setting, the cluster manager (CM), not the OS kernel, is best suited to make policy decisions.

General-purpose kernel code for making policy decisions is forced to handle all cases under unknown workloads, fostering implementation and performance complexity.
For example, in Linux fast-path failures are handled by complex fallback paths~\cite{haoMittOSSupportingMillisecond2017}.
At scale, they lead to performance anomalies that are hard to debug and harder to fix.
For example, several databases recommend disabling the kernel's automatic huge page promotion due to unpredictable latency spikes~\cite{couchbasethp,thpmongo,oracleinc.DatabaseInstallationGuide,RedisLatencyProblems}.

Meanwhile, leveraging the relative homogeneity of hardware and software in many WSCs can improve performance.
Thousands of identical tasks run across a WSC for hundreds of machines-years, and the workload mix changes incrementally as software teams update their services.
Yet the kernel treats each process as if it is the first and last of its kind.
It assumes little about new processes and uses the same stock policies for decisions.

To address kernel policy deficiencies and leverage WSC workload opportunities, we propose moving all policy decisions from the kernel to the CM, in a new paradigm we call \sys{}.
Each kernel monitors local system behavior, sending metrics to the CM.
The CM aggregates historical and cluster-wide metrics to make more optimal policy decisions, which it sends to the kernels.
The CM may also download a \textit{preset} into the kernel -- a limited policy for handling frequent or latency-sensitive decisions without a network round-trip.
Like software-defined networking~\cite{al-faresScalableCommodityData2008,kirkpatrickSoftwaredefinedNetworking2013}, where a central controller makes policy choices and individual switches use simple rules and tables, \sys{} benefits from global planning and simple, fast individual nodes.

\sys{} can use historic workload metrics to identify workloads suitable for eager memory allocation, resulting in a 15\% improvement in memcached response latency.
It can also automatically run experiments to identify the best set of pages to promote to huge pages.
Notably, our examples are low-hanging fruit; our design exposes opportunity for much improvement over the status quo.

\section{Target Setting}

\paragraph{Definitions.}
We define a \textit{warehouse-scale computer (WSC)} as a fairly homogeneous set of machines inter-connected via a high-speed network and running relatively stable, large-scale distributed systems.
A \textit{policy} is any kernel component that makes a runtime decision dynamically based on environmental inputs, including application behavior.
Examples include when to schedule processes or flush dirty blocks, whether to use huge pages, or whether to run a background thread such as memory compaction.
A \textit{mechanism} is an operation implemented in the kernel to accomplish some (usually hardware-related) objective.
Examples include context-switching, low-level I/O primitives, virtual memory mapping, or physical memory allocation.

\paragraph{System Model.}
We target settings in which (1) large amounts of cluster metrics can be aggregated and used over time, (2) communication within the WSC has a latency of dozens of microseconds or less, and (3) humans rarely interact with machines, and then only through a CM that automatically manages the life cycle and resource allocation of applications and machines.

Our target setting has relatively stable and homogeneous hardware and software.
Most machines in the cluster are similar to a large number of other machines (not necessarily a majority) in both hardware and software mix.
Major changes occur infrequently, but there may be frequent incremental software updates.
Many production WSCs satisfy these conditions\cite{microsoftinc.MicrosoftAzureTraces,googleinc.BorgClusterTraces,johnwilkesMoreGoogleCluster,tianCharacterizingSynthesizingTask2019,tirmaziBorgNextGeneration2020}.

\section{Kernel policies considered harmful}

In WSCs, built-in kernel policies beget many unnecessary kernel complexities and performance anomalies.
Also, abundant cluster metrics are available for policy design and execution, but they are not used to their full potential.

\begin{figure}
\includegraphics[width=\columnwidth]{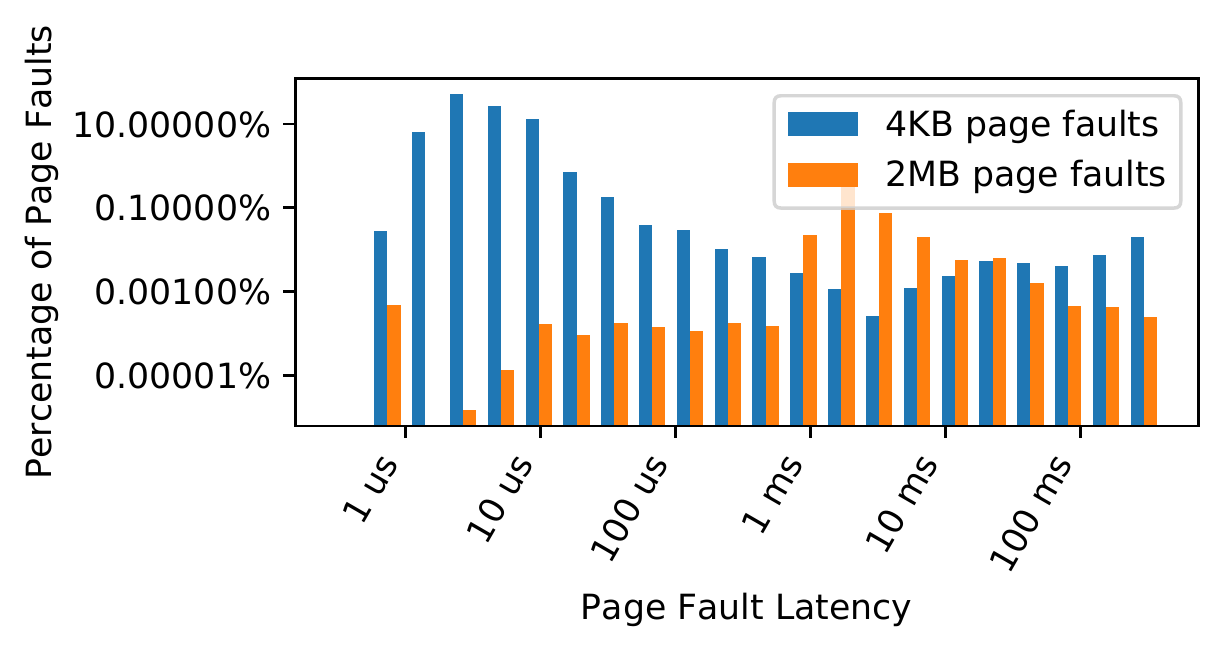}
\caption{Histogram of page fault latency for 350GB mix workload. Notice that both axes use log scales.\label{fig:pflatency}}
\end{figure}

\paragraph{Detrimental Complexity.}
Typically, machines in a WSC run a commodity kernel, such as Linux, with minimal modifications.
Commodity kernels include built-in policy implementations that are significantly complicated by the goal of generality.
But WSCs do not benefit from the added complexity.
For example, in Linux, the physical memory allocator slow-path is a jumble of policy and mechanism to anticipate all possible fast-path failures, such as low memory or high fragmentation.
The slow-path contains calls to the memory reclamation daemon, the memory compaction slow-path, and the synchronous memory reclamation slow-path.
It will try and retry each of these mechanisms to the extent allowed by the context of the allocation before resorting to the out-of-memory killer.
However, in a WSC, memory allocation and overcommitment are carefully controlled by operators; beyond simple fallback paths, allocation failures should raise an alert.

Moreover, generality harms performance.
Figure \ref{fig:pflatency} shows page fault latency on Linux for a 350GB workload.
Page fault latency varies over 6 orders of magnitude!
Linux must somehow decide whether to allocate a base or huge page, attempt huge page promotion, share a COW page (e.g., a zero page), reclaim memory, etc.
Often, its first choice fails and a fallback path executes, leading to high latency.
In contrast, prior work has found that failing fast allows more predictable WSC behavior~\cite{haoMittOSSupportingMillisecond2017}.

\begin{figure*}[th!]
\includegraphics[width=0.333\textwidth]{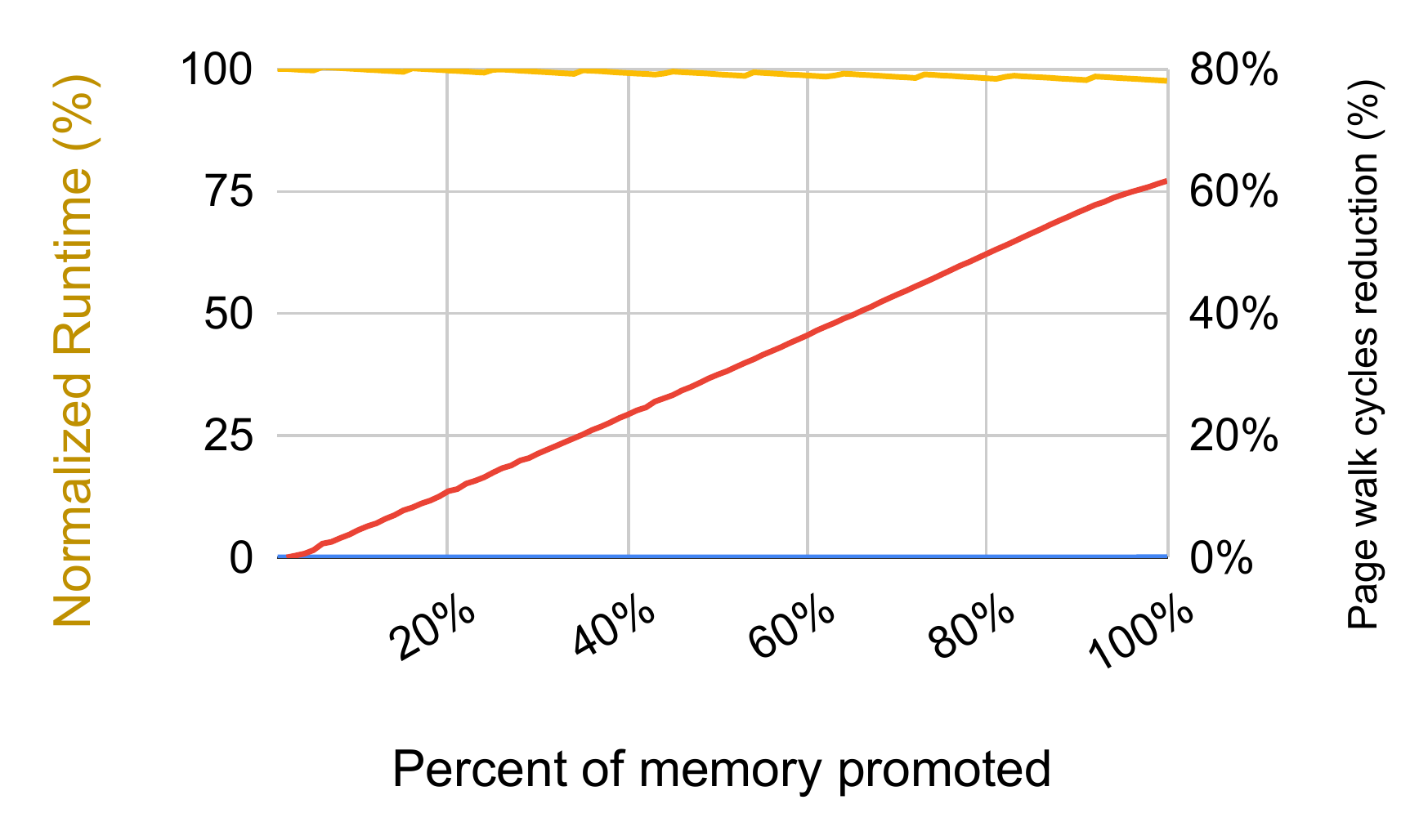}
\includegraphics[width=0.333\textwidth]{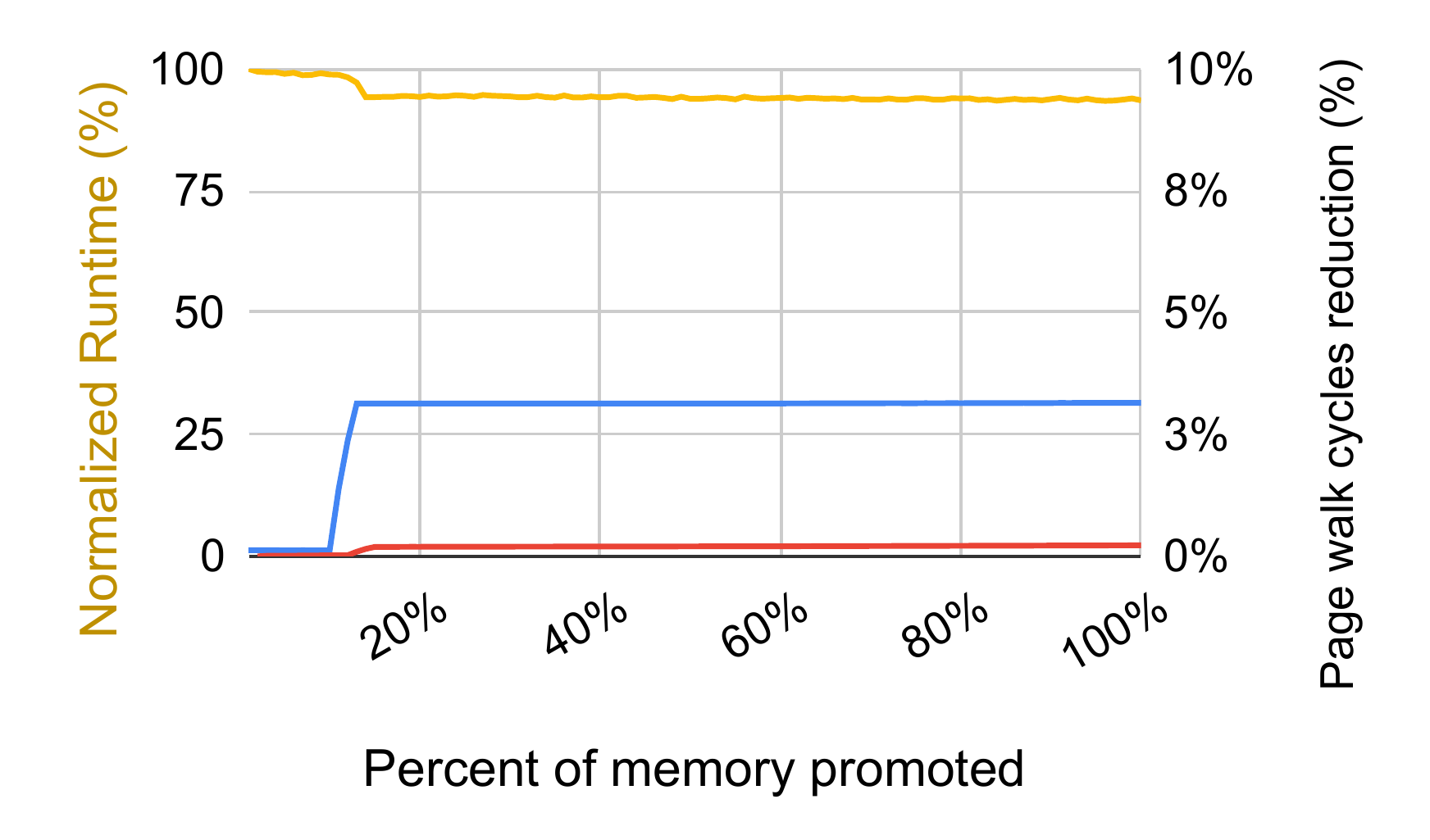}
\includegraphics[width=0.333\textwidth]{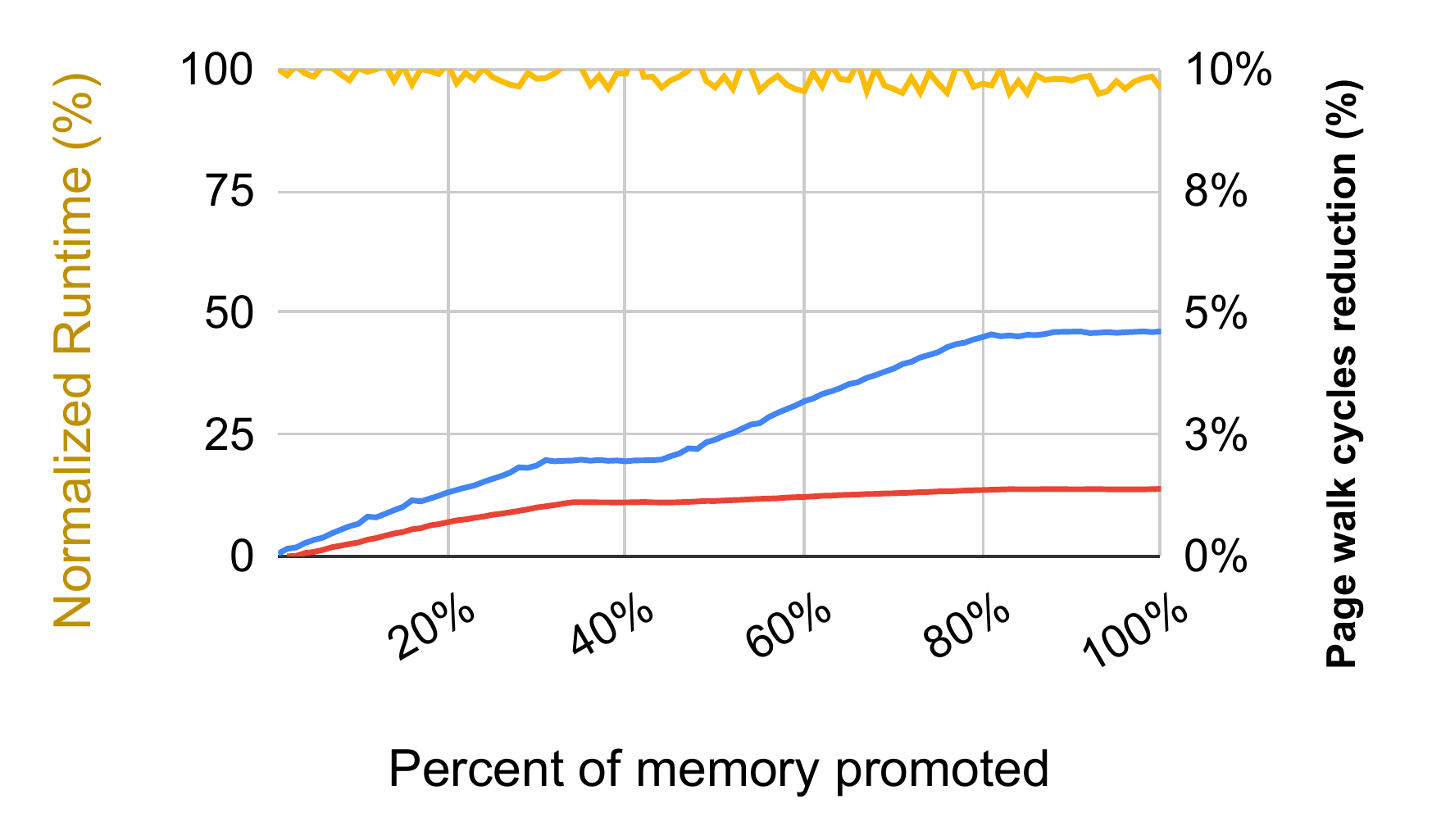}
\caption{Improvement in runtime (yellow), store page walks (red), and load page walks (blue) as more huge pages are used for \texttt{ubmk}, \texttt{xz}, and \texttt{mcf}, respectively. \label{fig:thp-benefit}}
\end{figure*}

\paragraph{Cluster Metrics and History.}
WSC workloads are ripe to be accurately and automatically characterized.
They are controlled and carefully allocated.
Changes occur gradually as software teams update their services.
The same applications run for thousands of machine-hours on the same machines continually~\cite{barrosoDatacenterComputerIntroduction2013,beyerSiteReliabilityEngineering2016,microsoftinc.MicrosoftAzureTraces,googleinc.BorgClusterTraces,johnwilkesMoreGoogleCluster,tianCharacterizingSynthesizingTask2019,tirmaziBorgNextGeneration2020}.

Cluster managers can experiment with policies on a subset of nodes and improve policies for all nodes.
For example, we measure the benefit of varying amounts of huge pages for three programs: a microbenchmark (\texttt{ubmk}) that allocates and writes to memory sequentially, and \texttt{xz} and \texttt{mcf} from SPEC (Figure \ref{fig:thp-benefit}).
\texttt{ubmk} sees up to 60\% reduction in page walks but bottlenecks on memory bandwidth, so runtime does not decrease.
\texttt{xz} sees 7\% improvement in runtime from promotion of a small number of pages.
\texttt{mcf} sees 5\% improvement in runtime from promotion of two large regions.
This characterization gives the precise benefit of promoting different pages in each workload and shows the optimal set of pages to promote given a budget.
The CM can generate this data by instructing different machines to map different sets of huge pages and aggregating the results.
Similarly, prior work has explored how to quantify performance and security isolation in clusters by aggregating data from many experiments~\cite{marsBubbleUpIncreasingUtilization2011,delimitrouBoltKnowWhat2017,romeroMageOnlineInterferenceaware2018}.

\begin{figure}
\includegraphics[width=\columnwidth]{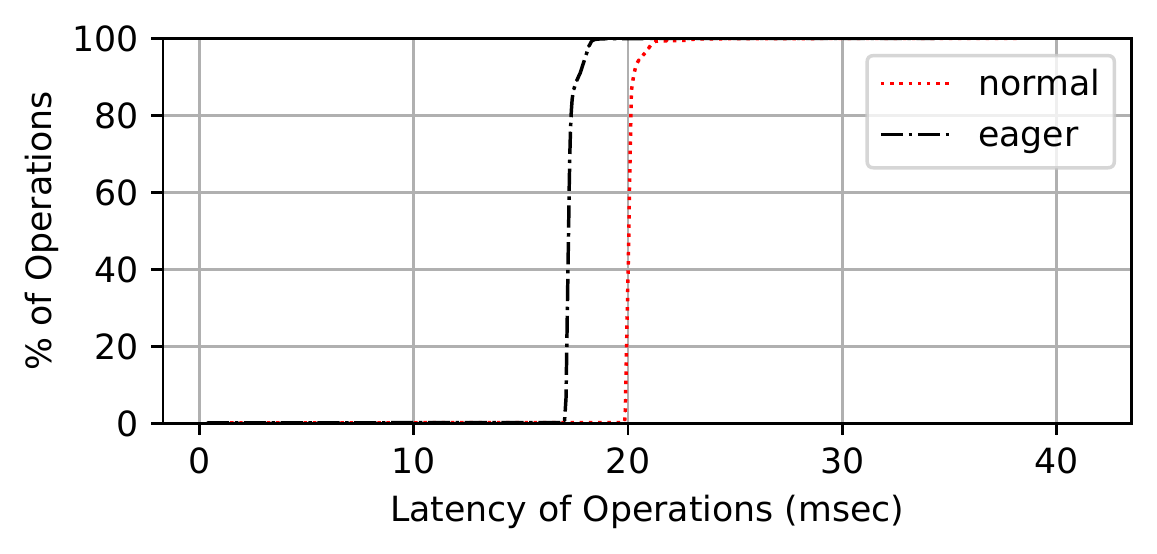}
\caption{64GB memcached workload with and without eager paging. 1 operation = 100 insertions. \label{fig:eager}}
\end{figure}

Historical data presents another major opportunity.
Commodity kernels assume little prior knowledge of a program, but in a WSC, completely new binaries are uncommon.
Rather, metrics from previous executions of a binary can inform policy for future executions.
For example, with \textit{eager paging}, the kernel eagerly allocates physical memory, rather than lazily on a page fault (the default)~\cite{karakostasRedundantMemoryMappings2015}.
If the process uses its entire allocation, eager paging avoids page faults during the workload.
Figure \ref{fig:eager} shows the CDF of operation latency for a memcached workload.
Eager paging improves latency by 3ms (15\%) for this workload without changing throughput or memory usage.
However, other workloads see up to 11\% longer latency for memory allocations or up to 125\% bloat in memory usage when using eager paging~\cite{karakostasRedundantMemoryMappings2015}.
The first time a process runs, the CM can passively monitor it to determine whether it would benefit from eager paging.
Subsequently, the CM can instruct all machines to use eager paging for this program.

\section{Look Ma! No kernel polices!}

\begin{figure}
\includegraphics[width=\columnwidth]{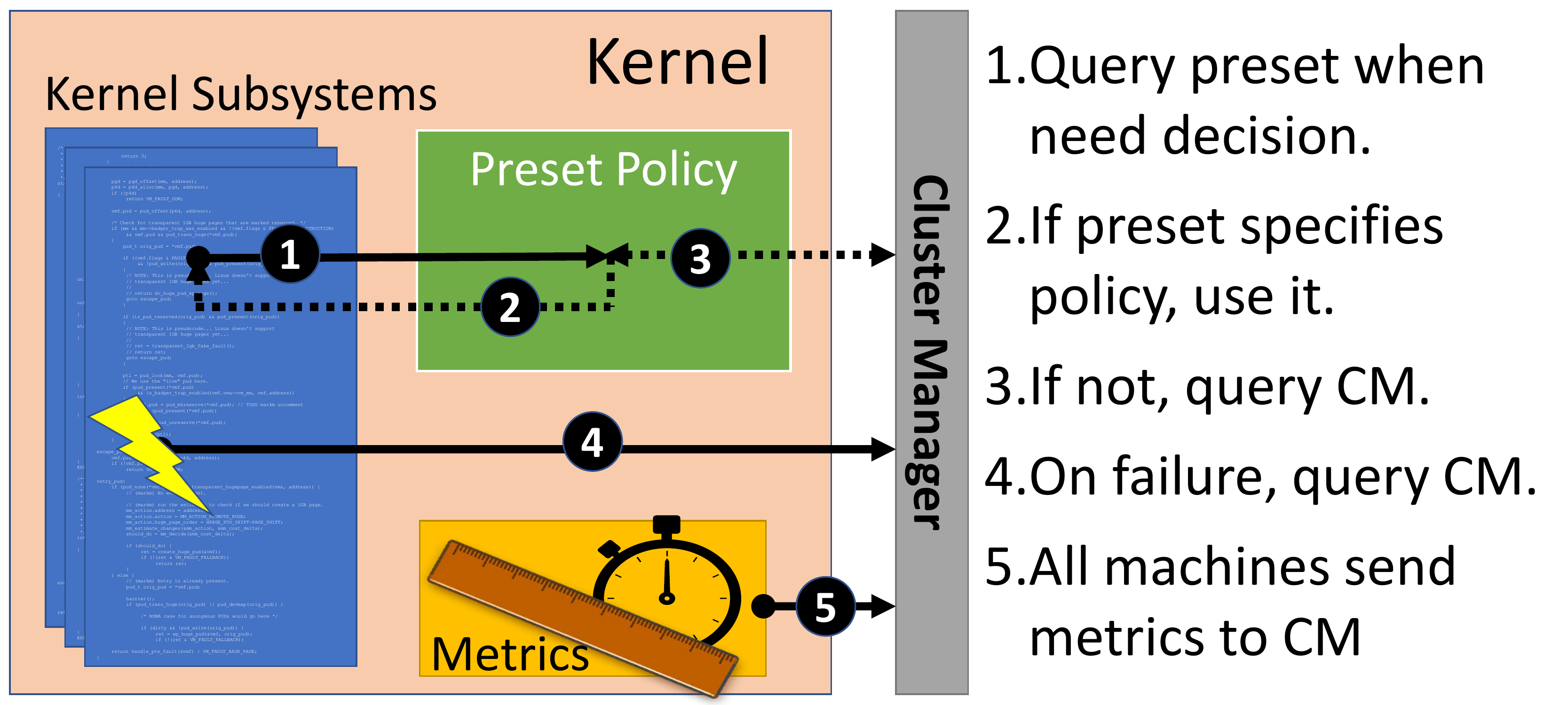}
\caption{Design Overview of \sys{}\label{fig:design}}
\end{figure}

Local kernel policies harm system performance, predictability, and implementation.
And, current systems do not take advantage of the relative homogeneity and stability of the WSC setting.
\textit{Our key claim is that the CM, not the kernel, should make all policy decisions by leveraging WSC metrics.}
Moving policy making out of the kernel simplifies system performance and implementation.
It also enables the CM to use cluster-wide and historical metrics to improve policy decisions.
Figure \ref{fig:design} shows an overview of our design, which we call \sys{}.

We implement a partial prototype which focuses on huge page policy.
In this section, we use huge page management as a running example of our proposal.

\subsection{From Kernel to CM}

\begin{lstlisting}[float=t,
caption={Sample policy query and response.}, label={fig:mmex2}]
# Sample query to cluster manager
{ type: "alloc-failure",
  process: "memcached",
  context: {"error": "page-fault-huge-page-alloc",
  	current-mem-usage: 103, #GB
	cpu-usage: 10, ...}                      }

# Sample response from cluster manager
{ action: "alloc-base-page",
  temporary-modify-preset:
    [{for: "1h", use-huge-pages: []},
     {for: "1h", mem-reclaim:
       {from: "my-low-priority-batch-job",
        addr: "0x8000000-0xf000000"}}]           }
\end{lstlisting}

\begin{lstlisting}[float=t,
caption={Sample preset policy.},
label={fig:mmex}]
{ mem-alloc-default: "demand-paging",
  mem-alloc-exceptions:
    {"memcached": "eager"},
  copy-on-write: "unspecified", # fall back to CM
  copy-on-write-exceptions:
    {"redis-snapshot": "no-cow"},
  page-size-default: 4096,
  use-huge-pages:
    [{"memcached": [0x435a0000, 0x435c0000, ...]},
     {"vid-encoder" [0x7ff000000, ...]}]
  numa-balancing: "local",
  out-of-memory: "unspecified", # fall back to CM
  mem-reclaim: "off",
  page-compaction:
    {when: "midnight", max-duration: "1s",
     max-cpu: 0.02},
  page-zeroing:
    {interval: "30s", max-cpu: 0.02},
  huge-page-promotion-async: "off",
  dirty-access-bit-scan:
    {interval: "30s", max-cpu: 0.10},            }
\end{lstlisting}

\begin{figure*}[t]
\footnotesize
\begin{tabular}{|clp{4.5in}|}
\hline
Type & Policy & How the CM can do it better \\ \hline
\multirow{4}{*}{Scheduling} & 
  Energy consumption &
  Increase safe power oversubscription~\cite{sakalkarDataCenterPower2020} \\
&
  CPU sleep states &
  Use WSC-wide load and service priorities to avoid slow-wakeup sleep states \\
&
  Co-location &
  Measure interference; isolate for performance and security~\cite{marsBubbleUpIncreasingUtilization2011} \\
&
  Thread affinity &
  Place threads that communicate on well-connected NUMA nodes~\cite{lepersThreadMemoryPlacement2015} \\
\hline
\multirow{3}{*}{Mem. Mgmt.} &
  Daemon CPU usage &
  Compute cost~\cite{buttFindingMoreDRAM2019,lagar-cavillaSoftwareDefinedFarMemory2019} and benefit of kswapd, kcompactd, khugepaged; disable when harmful \\
&
  NUMA placement &
  Place threads near their data based on long-latency memory access metrics~\cite{lepersThreadMemoryPlacement2015} \\
\hline
\multirow{3}{*}{I/O} &
  Buffer size &
  Determine best I/O buffer size for various applications \\
&
  Cache allocation &
  Determine best allocation of CPU and application caches based SLOs~\cite{bergerRobinHoodTailLatencyaware2018} \\
&
  Block I/O scheduling &
  Reduce device wear by coalescing writes; increase predictability~\cite{haoLinnOSPredictabilityUnpredictable2020} \\
\hline
\end{tabular}
\caption{Other example policies that benefit from the CM's scale and metrics collection.\label{fig:expol}}
\end{figure*}

In \sys{}, local kernels never make policy decisions independently; instead, they query the CM when a policy decision is needed.
Additionally, the CM may initiate a policy change (e.g., to move idle memory to far-memory, as in Google's far-memory system~\cite{lagar-cavillaSoftwareDefinedFarMemory2019}).
Listing \ref{fig:mmex2} exemplifies a query request and response after a huge page allocation failure.
In the example, the CM asks the kernel to stop allocating huge pages for some time and reclaim idle memory from a process.
Figure \ref{fig:expol} shows other policies and past work complemented by \sys{}.

In our implementation, we insert hooks at policy decision points in the kernel, allowing control through a \texttt{sysfs} interface.
For example, we add two hooks in the page fault handler and khugepaged.
We run hundreds of experiments sequentially to simulate gathering data from a large cluster (Figure \ref{fig:thp-benefit}).
Using this data, we select a subset of pages to promote.
On our system, mcf achieves 86\% of the benefit of THP with 42\% less internal fragmentation overhead due to huge pages.
Our implementation's simplicity allows greater insight into and control over the performance of the system.

\paragraph{Preset Policies.}
The local kernel must make some policy decisions when contacting the CM is impractical.
For example, scheduling and page fault handling are frequent and performance-critical; querying the CM each time would have massive performance and network overheads.
Thus, the CM downloads a \textit{preset policy} into the local kernel.
A preset is a policy that allows the kernel to make limited decisions without contacting the CM.
We do not specify what a preset policy looks like, but possible forms include a match-action table (like in an SDN~\cite{bosshartForwardingMetamorphosisFast2013}), an eBPF program, or an automaton.
Preset policies are limited; they do not handle edge cases or errors but fall back to the CM.
This keeps the policy simple and fast to execute.
It also informs the CM of exceptions, so it can improve the preset or alert an operator.

Listing \ref{fig:mmex} exemplifies a preset policy.
It specifies both default policy choices and actions to take in specific cases.
For example, on a page fault, the kernel checks if the faulting address is in \texttt{use-huge-pages}.
If it is, it attempts to allocate a huge page (otherwise, a base page).
If an error occurs, the kernel will query the CM.

Preset policies can improve performance over current systems.
For example, when Linux's page fault handler fails to allocate a huge page, it attempts page compaction or swapping, which can take dozens of milliseconds, often without fruit (Figure \ref{fig:pflatency}).
In contrast, preset policies fall back to the CM in uncommon cases, averting costly computation and long tail latency when it is wasteful.

\subsection{Policy Generation\label{sec:policydes}}

The CM makes policy decisions for all machines in the cluster using the metrics it collects from the cluster (see Section \ref{sec:collect}).
Policy decisions may apply cluster-wide (e.g., all machines should move 2GB to far-memory) or for specific machines (e.g., swap out a particular page).
We do not specify how the CM acquires policies, but many possibilities exist.
Google's far-memory system uses a Q-learning algorithm~\cite{lagar-cavillaSoftwareDefinedFarMemory2019}.
Other work suggests using neural networks~\cite{haoLinnOSPredictabilityUnpredictable2020}.
Our prototype uses a simple parameterized template that accepts a list of address ranges that have the highest impact when stored in huge pages.

The CM explores the space of policy decisions using a data-driven, partially-automated process.
Human experts provide a set of tunable kernel mechanisms and service-level objectives or performance goals.
As a workload runs, the CM tests different parameters across subsets of the cluster.
As data builds up, the CM uses statistical methods (possibly including machine learning) to find the best parameters under different conditions or to eliminate parts of the search space.
For example, our prototype measures the TLB-miss reduction of huge pages for large chunks of the address space and then narrows down to more promising regions.
Prior work also demonstrates the potential of CM-based policy exploration; an autotuner was able to increase far-memory efficacy by 5\% even after months of expert hand-tuning~\cite{lagar-cavillaSoftwareDefinedFarMemory2019}.

\subsection{Metrics Collection\label{sec:collect}}

The CM collects metrics from machines to make policy decisions for the whole cluster and individual machines.
Useful metrics may include amount of free, remote, idle, or fragmented memory, memory access patterns, per-process resource usage, core temperature, TLB/cache misses, IPC, and device performance.
Metrics collection must be efficient but frequent enough to detect changes in behavior (e.g., daily load variations).
As a baseline, if 10,000 machines send 100KB of metrics per second (e.g., 25,000 4-byte counters), the CM will receive merely 1GBps of metrics.
Other metrics may be too large to send frequently or may not need frequent reporting (e.g., memory usage data may be reported every 30s~\cite{lagar-cavillaSoftwareDefinedFarMemory2019}).
Moreover, stable metrics may be collected less frequently or from a subset of machines.
The change in each metric and allowable staleness are measured to inform the frequency of collection, as in~\cite{lagar-cavillaSoftwareDefinedFarMemory2019}.
Our experiments gather TLB miss data and other metrics once at workload termination -- roughly 1000B every 15 minutes.

\subsection{Discussion}

\paragraph{Coordination.}
Moving policies to the CM enables it to coordinate across machines to avoid bottlenecks in a distributed computation.
Google found that background activities, such as garbage collection, increase tail latencies because at any given time at least one machine is slow.
Cluster-wide coordination can eliminate this bottleneck~\cite{deanTailScale2013}.
Similarly, \sys{} enables coordinating kernel-level background tasks, such as memory compaction.

\paragraph{Practical Implementation.}
Unlike other kernel designs (e.g., unikernels~\cite{madhavapeddyUnikernelsLibraryOperating2013} or exokernels~\cite{englerExterminateAllOperating1995}), our design can be retrofitted into existing commodity kernels, such as Linux or Windows.
The relevant kernel code can be enabled or disabled using compile-time configuration.
Also, our proposal can be implemented incrementally by moving individual policies to the CM.
For example, Google moved far-memory management to the CM while leaving other memory management policies intact~\cite{lagar-cavillaSoftwareDefinedFarMemory2019}.
Thus, our proposal is compatible with high-availability requirements that make sweeping changes impossible.

\section{Related Work}

There has been significant work on cluster management and scheduling~\cite{burnsBorgOmegaKubernetes2016,hindmanMesosPlatformFinegrained2011,kennyyuTupperwareEfficientReliable2019,vavilapalliApacheHadoopYARN2013,vermaLargescaleClusterManagement2015,kennyyuTupperwareEfficientReliable2019}.
Prior work uses cluster-wide metrics and policies to improve efficiency~\cite{lagar-cavillaSoftwareDefinedFarMemory2019,tirmaziBorgNextGeneration2020,rzadcaAutopilotWorkloadAutoscaling2020} and performance isolation~\cite{marsBubbleUpIncreasingUtilization2011,delimitrouBoltKnowWhat2017,romeroMageOnlineInterferenceaware2018} in WSCs.
\sys{} goes further by suggesting that all kernel policies should move to the CM.
Our work can leverage prior work to identify sources of performance unpredictability in WSCs~\cite{ganSeerLeveragingBig2019,yanqizhangSinanMLBasedQoSAware,yuganSagePracticalScalable}

Software-defined networks are the networking analogue of \sys{}: a global network controller sets policies, while individual switches route traffic based on simple tables.
This leads to higher performance and flexibility built with simple switches~\cite{al-faresScalableCommodityData2008,kirkpatrickSoftwaredefinedNetworking2013,koponenNetworkVirtualizationMultitenant2014}.

Improving policy decisions requires accurate and precise data~\cite{bangaResourceContainersNew1999,kwonCoordinatedEfficientHuge2016,panwarHawkEyeEfficientFinegrained2019,balakrishnanIntegratedCongestionManagement1999}, but it can be expensive to collect.
Google and Facebook report a cost of 10\% CPU for memory usage metrics~\cite{buttFindingMoreDRAM2019}.
\sys{} amortizes costs over the cluster and makes more efficient use of large-scale deployments for metrics gathering.

Cluster-wide workload traces have been published~\cite{microsoftinc.MicrosoftAzureTraces,googleinc.BorgClusterTraces,johnwilkesMoreGoogleCluster,tianCharacterizingSynthesizingTask2019,tirmaziBorgNextGeneration2020} and used to reduce memory fragmentation~\cite{maasLearningbasedMemoryAllocation2020}.
Other work uses live profiling data on individual machines to improve fragmentation~\cite{chenProfileguidedProactiveGarbage2006}, I/O scheduling~\cite{haoLinnOSPredictabilityUnpredictable2020}, and NUMA placement~\cite{lepersThreadMemoryPlacement2015}.
Our work complements and extends this work by using cluster-wide live metrics to make policy decisions for all nodes.

There has been much prior work on the structure of OS kernels.
Hydra proposed separating policy from mechanism, and included a similar mechanism to our preset policies~\cite{hydra}.
Unikernels hard-code policies into a library OS linked to the application~\cite{madhavapeddyUnikernelsLibraryOperating2013}.
Exokernels move policy decisions into userspace~\cite{englerExterminateAllOperating1995}.
These approaches are complementary to \sys{}, providing a way to move policies out of the kernel.

\section{Conclusion}

We propose moving all policy making into the CM and removing it entirely from the OS kernel.
This leads to better decision-making across the cluster by taking advantage of ample cluster-wide and historical workload metrics and a relatively constant workload mix.

By aggregating cluster-wide profiles, the CM is able not only to make better decisions itself, but to give operators greater visibility into their systems.
We believe this will open doors for future optimizations that are currently impossible to implement.
We also believe it will simplify system behavior and kernel implementation significantly.

\section*{Acknowledgements}

We thank our colleagues and anonymous reviewers for helpful feedback on our work.

This work was funded by NSF grants CNS 1815656 and CNS 1900758.

\bibliography{references}{}
\bibliographystyle{plain}

\end{document}